\documentclass[a4paper,11pt]{article}
\usepackage[utf8]{inputenc}
\usepackage{graphicx}
\usepackage{float}
\usepackage{wrapfig}
\usepackage{multirow}
\usepackage{hhline}
\usepackage{setspace}
\usepackage[labelfont={small,bf},textfont={small}]{caption}
\usepackage{subcaption}
\usepackage[dvipsnames,table]{xcolor}
\usepackage[linktocpage=true,colorlinks=true,
            linkcolor=NavyBlue,citecolor=NavyBlue,
            urlcolor=Mahogany]{hyperref}
\usepackage{amsmath,amssymb,amsfonts}
\usepackage{kpfonts}
\usepackage{anyfontsize}
\usepackage{appendix}
\usepackage{geometry}
\usepackage{braket}
\usepackage{cite}

\def\eqref#1{\textcolor{NavyBlue}{(\ref{#1})}}
\newgeometry{top=3cm, bottom=3cm, right=2.5cm, left=2.5cm}

\begin{document}

%%%----------------------------------------------------------------------------
\begin{center}
{\Large\bfseries
Robust isolated attosecond pulse generation with self-compressed sub-cycle drivers from hollow capillary fibers\par}
\vspace{2ex}
{\large Marina Fernández Galán,$^{1,2,*}$ Javier Serrano,$^{1,2}$ Enrique Conejero Jarque,$^{1,2}$ Rocío Borrego-Varillas,$^3$ Matteo Lucchini,$^{3,4}$ Maurizio Reduzzi,$^4$ Mauro Nisoli,$^{3,4}$ Christian Brahms,$^5$ John C. Travers,$^5$ Carlos Hernández-García,$^{1,2}$ and Julio San Roman$^{1,2}$\par}
\vspace{2ex}
{\footnotesize $^1$Grupo de Investigación en Aplicaciones del Láser y Fotónica, Departamento de Física Aplicada, Universidad de Salamanca, E-37008 Salamanca, Spain \\ $^2$Unidad de Excelencia en Luz y Materia Estructuradas (LUMES), Universidad de Salamanca, Salamanca, Spain \\ $^3$Institute for Photonics and Nanotechnologies (IFN), Consiglio Nazionale delle Ricerche (CNR), Piazza Leonardo da Vinci 32, 20133 Milano, Italy \\ $^4$Department of Physics, Politecnico di Milano, Piazza Leonardo da Vinci 32, 20133 Milano, Italy \\ $^5$School of Engineering and Physical Sciences, Heriot-Watt University, Edinburgh, United Kingdom\par}
\vspace{2ex}
\textcolor{NavyBlue}{$^*$marinafergal@usal.es}
\vspace{2ex} 
\end{center}

%%%----------------------------------------------------------------------------
\begin{abstract}
High-order harmonic generation (HHG) arising from the non-perturbative interaction of intense light fields with matter constitutes a well-established tabletop source of coherent extreme-ultraviolet and soft X-ray radiation, which is typically emitted as attosecond pulse trains. However, ultrafast applications increasingly demand isolated attosecond pulses (IAPs), which offer great promise for advancing precision control of electron dynamics. Yet, the direct generation of IAPs typically requires the synthesis of near-single-cycle intense driving fields, which is technologically challenging. In this work, we theoretically demonstrate a novel scheme for the straightforward and compact generation of IAPs from multi-cycle infrared drivers using hollow capillary fibers (HCFs). Starting from a standard, intense multi-cycle infrared pulse, a light transient is generated by extreme soliton self-compression in a HCF with decreasing pressure, and is subsequently used to drive HHG in a gas target. Owing to the sub-cycle confinement of the HHG process, high-contrast IAPs are continuously emitted almost independently of the carrier-envelope phase (CEP) of the optimally self-compressed drivers. This results in a CEP-robust scheme which is also stable under macroscopic propagation of the high harmonics in a gas target. Our results open the way to a new generation of integrated all-fiber IAP sources, overcoming the efficiency limitations of usual gating techniques for multi-cycle drivers.  
\end{abstract}

%%%----------------------------------------------------------------------------
\section{Introduction}

Laser sources delivering broadband ultrashort pulses are of paramount importance for ultrafast science. In a continuous effort to access the briefest phenomena in nature, the temporal resolution afforded by this technology has advanced through twelve orders of magnitude in the last five decades, overcoming the ultimate limit set by the period of the carrier wave. In the last years, the generation of sub-cycle optical waveforms has enabled unprecedented control of electron dynamics and strong field processes \cite{Wirth2011,Hassan2016}. Among the latter, high harmonic generation (HHG) stands out as the only tabletop process capable of providing coherent extreme-ultraviolet (EUV) and soft X-ray radiation \cite{Popmintchev2012}. In the semi-classical microscopic picture of HHG in gaseous media \cite{Schafer1993}, an intense infrared (IR) laser pulse first tunnel-ionizes the atoms and coherently drives the motion of the free electrons. Second, after reversal of the driving electric field, the electrons are driven back to the parent ions and, upon recollision, high frequency radiation is emitted. As the entire HHG process is repeated every half cycle of the IR pulse, standard high harmonic emission consists of a train of attosecond bursts \cite{Farkas1992,Paul2001}. This unique ultrafast source has opened the door to time-resolved studies of valence electron motion in atoms \cite{Baltuska2003,Goulielmakis2010} or charge migration in molecules \cite{Calegari2014,Kraus2015,Nisoli2017}, among many others \cite{Shi2020,Borrego-Varillas2022}. Nevertheless, for certain applications, the isolation of a single attosecond pulse from the train is preferred. For this purpose, different gating techniques have been developed, allowing for the generation of isolated attosecond pulses (IAPs) from commercially-available multi-cycle IR pulses. These consist on controlling the rescattering process on the microscopic level like polarization gating \cite{Sansone2006,Sola2006,Sansone2009} or two-color and double optical gating \cite{Mashiko2008,Takahashi2010,Zhao2012}, taking advantage of macroscopic propagation effects like ionization gating or time-gated phase-matching \cite{Sandhu2006,Popmintchev2009,Thomann2009,Ferrari2010,Chen2014,Hernandez-Garcia2017}, or implementing the attosecond lighthouse effect based on spatio-temporal wavefront control \cite{Vincenti2012,Kim2013}.

Simpler techniques like amplitude gating with few-cycle drivers have also been demonstrated \cite{Christov1997,Goulielmakis2008}, but these require additional spectral selection of the high-energy cut-off produced by the most intense half cycle of the driving field, precluding the generation of ultrabroadband IAPs \cite{Chini2014}. Since the first experimental confirmation of amplitude gating in 2001 \cite{Hentschel2001}, this technique has been refined to overcome this bandwidth limitation by the use of ever shorter IR pulses down into the sub-cycle regime. These waveforms naturally confine the HHG process to the only intense half cycle electric field and, recently, precise tailoring of driving transients has allowed the direct creation of highly tunable IAPs and enhanced HHG spectra \cite{Wei2013,Jin2014,Yang2021}. However, sub-cycle control of light transients often requires the use of extremely complex systems like the so called parametric waveform synthesizers \cite{Rossi2020}. Therefore, next-generation HHG experiments would strongly benefit from the availability of more compact and handy sources of sub-cycle optical drivers.

A very promising alternative for the generation of sub-cycle IR pulses comes through high-energy soliton dynamics in gas-filled hollow capillary fibers (HCFs) \cite{Travers2019}. These simple fibers are routinely used for ultrashort pulse compression \cite{Nisoli1996}, and allow for significant energy scaling and nonlinearity and dispersion tuning by modifying the pressure of the filling gas \cite{Grigorova2023}. In particular, if the latter is chosen so that an input multi-cycle pulse propagates in the HCF with anomalous dispersion, the simultaneous nonlinear spectral broadening by self-phase modulation (SPM) and phase compensation arising from the negative group-velocity dispersion (GVD) can lead to soliton self-compression well down into the sub-cycle regime \cite{Voronin2014,Brahms2020}. Recent studies have demonstrated that this extreme pulse compression can be further enhanced by pumping the fiber with a decreasing pressure gradient \cite{Galan2022}, and that broadly similar high-quality sub-cycle IR fields can be generated in different HCF scenarios \cite{Galan2023}. In addition, the use of decreasing pressure could be of great interest for HHG experiments, as it allows for the direct delivery to vacuum of the self-compressed pulses free of distortions from transmission optics \cite{Brahms2019,BrahmsRDW2020}. Altogether, the combination of HHG beamlines with HCFs delivering intense self-compressed IR transients opens a very promising scenario to develop compact and versatile scientific tools for the generation of high-frequency IAPs, but theoretical investigations and design guidelines are still missing to make it a feasible technique.

In this work, we demonstrate a unique control in the efficient generation of IAPs from self-compressed multi-cycle IR pulses, combining for the first time in a compact scheme the advantages of using state-of-the-art femtosecond pump pulses with the possibilities offered by sub-cycle waveform tailoring. We numerically study HHG driven by IR sub-cycle pulses generated by extreme soliton self-compression in a gas-filled HCF with a decreasing pressure gradient which allows direct delivery to a vacuum beamline. By systematically scanning the energy of the input pulse to the fiber, its output carrier-envelope phase (CEP) and the pumping gas pressure, different IR fields are synthesized. When driving HHG with these unique waveforms, we can control the properties of the generated high-order harmonics and attosecond pulses. In particular, our results demonstrate that high-contrast IAPs are directly produced for a broad set of driving fields corresponding to the optimally self-compressed IR pulses. Most interestingly, owing to the nature of the IR waveforms, clean IAPs are continuously emitted for a wide range of driver's CEP, resulting in a CEP-robust scheme which is also stable under macroscopic propagation of the high harmonics in a gas target.

%%%----------------------------------------------------------------------------
\section{Methods}

\begin{figure}[htbp]
    \centering
    \includegraphics[width=0.85\linewidth]{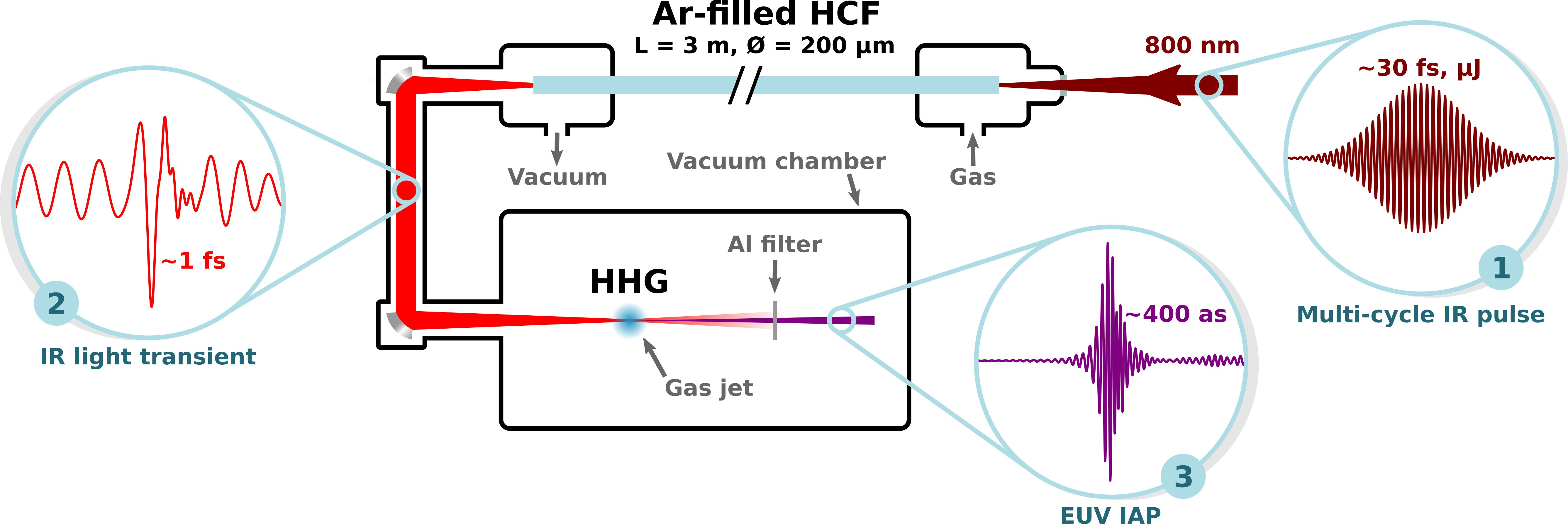}
    \caption{Schematic of an in-vacuum HHG beamline driven by sub-cycle self-compressed pulses from a gas-filled HCF with a decreasing pressure gradient. Starting from a multi-cycle IR pulse (1), a light transient (2) is generated by extreme soliton self-compression in the fiber and subsequently used to produce EUV IAPs (3) in a gas target.}
    \label{fig:f1_setup}
\end{figure}

Figure~\ref{fig:f1_setup} shows the proposed scheme comprising a first stage of pulse self-compression in a gas-filled HCF and subsequent HHG in a gas target driven by the sub-cycle waveforms exiting the fiber. The capillary is negatively pumped, i.e., filled with a decreasing pressure gradient, with the gas supplied at the entrance at a pressure $p_0$, and the output end directly coupled to the vacuum beamline at pressure $p_L=0$. In this way, the resulting pressure distribution along the longitudinal coordinate $z$ of the fiber is given by \cite{Suda2005}:
\begin{equation}
    p(z) = \sqrt{p_0^2+\frac{z}{L}\left( p_L^2-p_0^2 \right)} = p_0\sqrt{1-\frac{z}{L}}, \label{eq:gradient}
\end{equation}
where $L$ represents the HCF length. The first fiber stage is numerically modeled with a (2+1)D multi-mode propagation equation for the pulse complex envelope $A(\rho,T,z)$, including the complete dispersion and loss of the gas-filled HCF, self-phase modulation (SPM), self-steepening, and photoionization and plasma absorption \cite{Jarque2018,Crego2023}. In this (2+1)D model, the pulse propagation equation is integrated in a local time $T$ measured in a reference frame traveling with the pulse at the group velocity, and the spatial pulse envelope is assumed to have cylindrical symmetry and depend only upon the radial coordinate $\rho$. To first explore different self-compression scenarios and perform systematic parameter scans, this computationally-demanding model can be reduced to a time-dependent (1+1)D nonlinear propagation equation for the fundamental EH$_{11}$ mode of the HCF, neglecting spatial and plasma dynamics \cite{Crego2023}. This approximation accurately describes ultrashort pulse propagation in the low-intensity regime, where the peak power and the peak intensity of the pulse remain, respectively, below the critical power for self-focusing and the threshold intensity for gas ionization \cite{Nurhuda2003}.

When complete (2+1)D simulations are performed, the cylindrically symmetric pulse at the HCF output $A(\rho,T,z)$ is then propagated in vacuum and focused onto a low density gas target to drive HHG, as depicted in Fig.~\ref{fig:f1_setup}. The beam free-space expansion from $z$ to $z+\Delta z$ is numerically accomplished by expressing $A(\rho,T,z)$ as a superposition of plane monochromatic waves and multiplying each Fourier component by its corresponding propagation phase \cite{Couairon2011}:
\begin{gather}
    \Tilde{A}(k_{\rho},\omega,z) = \int_{0}^{\infty} \rho A(\rho,\omega,z)J_0(k_{\rho}\rho) \, d\rho , \\
    \Tilde{A}(k_{\rho},\omega,z+\Delta z) = \Tilde{A}(k_{\rho},\omega,z) \exp \left\lbrace i\Delta z \left( k_z(\omega)-k_{z,0}-\frac{\omega-\omega_0}{v_{g,0}} \right) \right\rbrace ,
\end{gather}
where $\rho = \sqrt{x^2+y^2}$ is the radial position in cylindrical coordinates, $\omega$ stands for angular frequency, $k_{\rho}$ and $k_z$ are radial and axial wave numbers, $\Tilde{A}(k_{\rho},\omega,z)$ is the Hankel transform of $A(\rho,\omega,z)$, $J_0(x)$ is the zero-th order Bessel function of the first kind, $A(\rho,\omega,z) = \mathcal{F}[A(\rho,T,z)]$, and $\mathcal{F}$ stands for direct Fourier transform. The dispersion relation in vacuum is given by $k_z^2 = \omega^2/c^2-k_{\rho}^2$, $c$ is the speed of light, $k_{z,0} = k_z(\omega=\omega_0,k_{\rho}=0)=\omega_0/c$ is the wave number of a plane wave at the central frequency $\omega_0$ propagating in the longitudinal direction, and $v_{g,0} = v_g(\omega=\omega_0,k_{\rho}=0) = c$ is the speed of a reference frame moving with the group velocity of the pulse, with $v_g = (\partial k_z/\partial\omega)^{-1}$. The beam is focused with an ideal concave mirror, which is modeled in the frequency domain as a quadratic spatial phase $\sim\exp\lbrace-i\omega \rho^2/(2cf)\rbrace$, $f$ being its focal length. This procedure recovers a perfect image of the HCF output at the focal plane \cite{Kaplan1998}, with a suitable intensity for driving efficient HHG.

The corresponding IR electric field around the focal volume is generated by adding the carrier wave as $E(\rho,T,z) = \Re[ A(\rho,T,z)\exp\lbrace -i\omega_0 T + i\phi_{\mathrm{CEP}} \rbrace ]$, $\phi_{\mathrm{CEP}}$ being the CEP, and it is used as input to macroscopic HHG calculations in a gas target. Note that, as $A(\rho,T,z)$ is a complex quantity with its own temporal phase, $\phi_{\mathrm{CEP}}=0$ might not correspond to the situation of maximum field amplitude. In the HHG simulations, the generation medium is discretized into elementary radiators and single-atom harmonic contributions are computed through the full integration of the three-dimensional time-dependent Schr\"odinger equation (3D-TDSE), under the single-active electron approximation. The resulting emissions are then propagated to a far-field detector through the electromagnetic field propagator \cite{Hernandez-Garcia2010}, thus taking into account the phase-matching of the high-order harmonics. The fundamental and low-order harmonics are finally filtered with a 200-nm-thick aluminum foil. Note that we considered that, after the HCF, the IR driving field propagates in vacuum, neglecting dispersion and nonlinear reshaping in the gas target, an assumption that is valid in the case of moderate pulse intensities and low density targets ($\sim 10^{17}$ atoms/cm$^3$) such as those used in standard HHG experiments \cite{Hernandez-Garcia2010}.

%%%----------------------------------------------------------------------------
\section{Results and discussion}

\subsection{Optimal sub-cycle self-compression and HHG regimes} \label{sec:scaling}

First, to produce a wide variety of ultrashort IR waveforms and investigate their influence on HHG, we have followed the procedure detailed in \cite{Galan2022}. In short, using the (1+1)D propagation model, we have simulated the self-compression of a 30-fs (intensity full width at half-maximum, FWHM) transform-limited gaussian pulse at 800 nm through a 3-m-long, 100-\textmu m-core-radius HCF filled with argon, while varying its initial energy $U_0$ and the equivalent pressure $p_{\mathrm{eq}}$. The latter is defined as the constant gas pressure which matches the nonlinear phase-shift acquired by the pulse during its propagation through the negatively pumped fiber and, for a pressure distribution in the form of Eq.~\eqref{eq:gradient}, it is simply related to the pumping pressure by $p_{\mathrm{eq}}=2p_0/3$ \cite{BrahmsRDW2020}.

\begin{figure}[b!]
    \centering
    \includegraphics[width=0.95\linewidth]{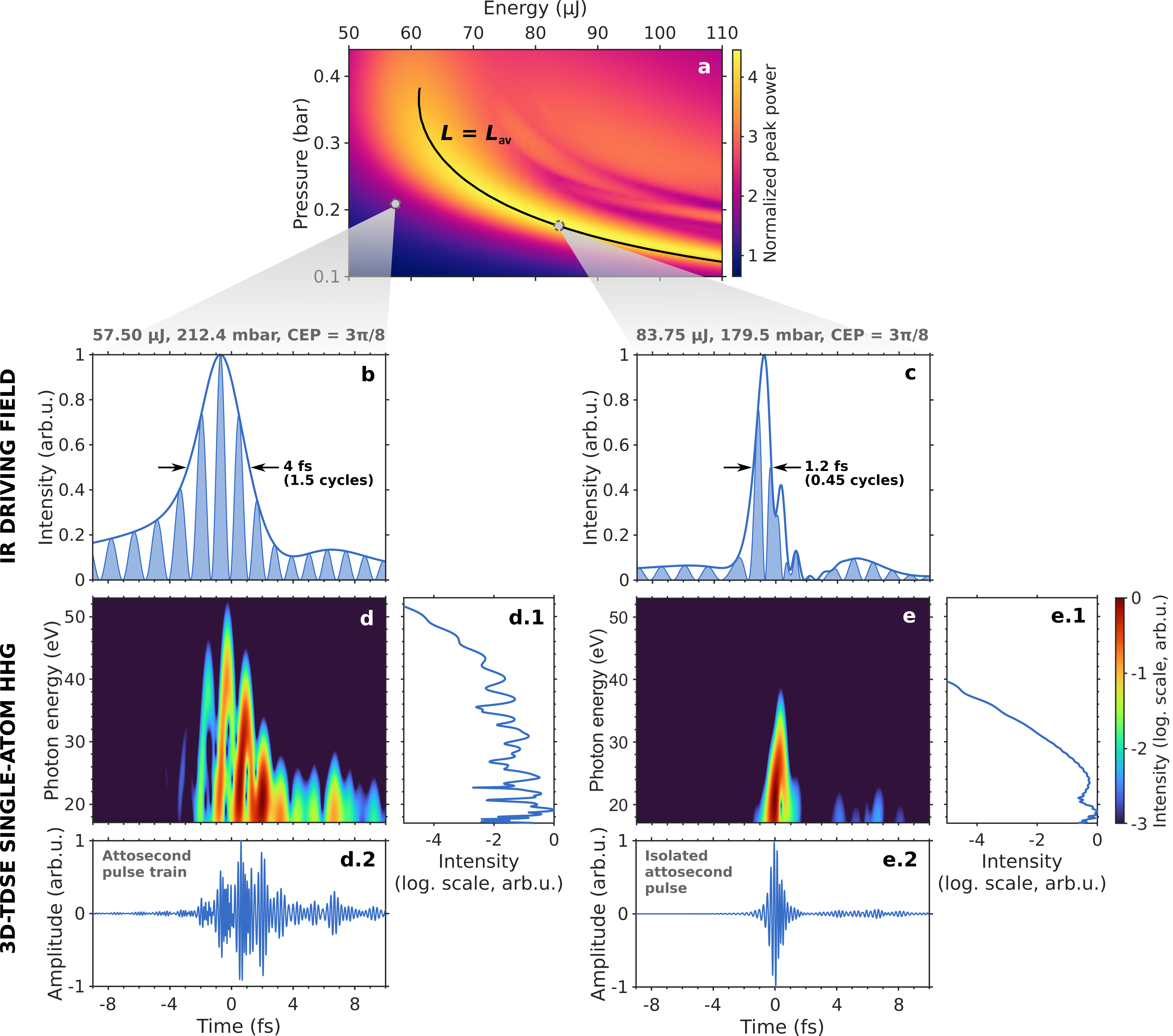}
    \caption{\textbf{(a)} Ratio of output to input peak power of the self-compressed IR pulses as a function of the initial energy $U_0$ and the equivalent constant pressure $p_{\mathrm{eq}}$ in the negatively pumped 100-\textmu m-core-radius, 3-m-long HCF filled with Ar. The solid black line represents the contour where $L=L_{\mathrm{av}}$, which runs along the optimal region for sub-cycle self-compression. \textbf{(b,c)} Self-compressed IR drivers generated in the HCF for two different sets of $(U_0,\,p_{\mathrm{eq}},\,\phi_{\mathrm{CEP}})$: \textbf{(b)} a few-cycle pulse and \textbf{(c)} a sub-cycle light transient. \textbf{(d,e)} Time-frequency analysis of their corresponding single-atom HHG emissions in hydrogen when their instantaneous peak intensity is set to $1.57 \times 10^{14}$ W/cm$^2$ (electric field amplitude of 0.067 au). Additional panels (d.1,e.1) show the high-harmonic spectra and (d.2,e.2) their temporal counterpart in the form of an attosecond pulse train or a clean IAP.}
    \label{fig:f2_timefreq}
\end{figure}

For each $(U_0,\,p_{\mathrm{eq}})$ pair, Fig.~\ref{fig:f2_timefreq}\textcolor{NavyBlue}{(a)} shows the ratio of output to input peak power of the self-compressed pulses. In this plot, the optimal region for high-quality sub-cycle pulse generation can be readily identified as the area of largest peak power enhancement, which was also found to overlap with the region of shortest output pulse duration \cite{Galan2022,Galan2023}. In the decreasing pressure configuration, this optimal self-compression region is in general delimited by two constraints. On one hand, the soliton order must be kept $N<15$ to achieve a high-quality compression without triggering modulation instabilities \cite{Voronin2008,Travers2011}. On the other hand, the fixed fiber length has to match the characteristic length of the process to ensure that the self-compressing pulse reaches the minimum possible duration without entering in the soliton fission regime. In previous works, an average self-compression length was defined as \cite{Galan2022}:
\begin{equation}
    L_{\mathrm{av}}=\frac{L_{\mathrm{sc}}+L_{\mathrm{fiss}}}{2}=\frac{1+\sqrt{2}}{2\sqrt{2}}L_{\mathrm{fiss}} ,
\end{equation}
where $L_{\mathrm{fiss}}=L_D/N$ is the fission length \cite{Travers2011}, $L_{\mathrm{sc}}=L_{\mathrm{fiss}}/\sqrt{2}$ is a self-compression length \cite{Chen2002}, $N=(L_D/L_{NL})^{1/2}$ is the soliton order, and $L_D=T_p^2/(4\ln{2}|\beta_2|)$ and $L_{NL}=1/(\gamma P_0)$ determine the characteristic length scales of GVD and SPM, respectively \cite{Agrawal}. Here $T_p$ represents the FWHM duration of the input gaussian pulse, $P_0$ refers to its peak power, $\beta_2$ is the GVD coefficient of the HCF and $\gamma$ is the nonlinear parameter as defined elsewhere \cite{Agrawal}. As also shown in Fig.~\ref{fig:f2_timefreq}\textcolor{NavyBlue}{(a)}, the condition $L=L_{\mathrm{av}}$ describes a contour line in the energy-pressure plane which, when falling inside the space with $N<15$, can be used to identify the optimal region for high-quality self-compression in any configuration \cite{Galan2023}.

After simulating the first HCF stage, we have generated different IR fields from each self-compressed pulse by adding the carrier wave with eight values of $\phi_{\mathrm{CEP}}$ ranging from $0$ to $\pi$~rad. We verified with a carrier-resolved propagation code that, in our range, this can be equally accomplished by varying the CEP of the input pulse, which is a more realistic experimental scenario. However, from the computational point of view, we found it more efficient for the parameter sweeps to propagate the pulse envelope and then build the electric field at the HHG generation points. The resulting large set of waveforms, each identified by $(U_0,\,p_{\mathrm{eq}},\,\phi_{\mathrm{CEP}})$, was used to perform single-atom 3D-TDSE HHG calculations in atomic hydrogen. This target species was chosen for simplicity, but the 3D-TDSE could be easily extended to other noble gases (Ar, He, Kr, Ne), under the single-active electron approximation. In addition, to isolate the influence of the driving waveform itself in HHG, all IR drivers were first normalized to an instantaneous peak intensity of $1.57 \times 10^{14}$ W/cm$^2$ (corresponding to an electric field amplitude of 0.067 atomic units, au). Experimentally, this could be achieved by inserting a variable attenuator in the beam path in Fig.~\ref{fig:f1_setup} or by adjusting the focusing geometry. Finally, the temporal and spectral properties of the harmonic radiation (maximum photon energy, isolation of attosecond pulses, contrast, etc.) were analyzed in terms of the free parameters $(U_0,\,p_{\mathrm{eq}},\,\phi_{\mathrm{CEP}})$, providing useful design guidelines for experiments.

Two representative examples of HHG driven by IR self-compressed pulses are shown in Figs.~\ref{fig:f2_timefreq}\textcolor{NavyBlue}{(b-e)}. In the region below the contour line $L=L_{\mathrm{av}}$, the propagating pulse exits the fiber before reaching the maximum self-compression point, resulting in driving fields with few-cycle durations like the one shown in Fig.~\ref{fig:f2_timefreq}\textcolor{NavyBlue}{(b)}. The results of the 3D-TDSE HHG calculations for this IR pulse are shown in the time-frequency analysis (also known as spectrogram) in Fig.~\ref{fig:f2_timefreq}\textcolor{NavyBlue}{(d)}, which encodes the complete information of the HHG emission both in intensity and phase. As we can see in panel~\ref{fig:f2_timefreq}\textcolor{NavyBlue}{(d.2)}, this kind of sub-optimal self-compressed pulses yield attosecond pulse trains in the temporal domain, but their HHG spectra (panel~\ref{fig:f2_timefreq}\textcolor{NavyBlue}{(d.1)}) were found to reach the highest photon energies because the field strength is preserved after the ionization HHG step with longer pulses.

As the product $U_0 \times p_{\mathrm{eq}}$ is increased, the output pulses from the HCF become shorter until high-quality sub-cycle waveforms are generated in the region around the contour $L=L_{\mathrm{av}}$. Figure~\ref{fig:f2_timefreq}\textcolor{NavyBlue}{(c)} shows one of such IR light transients, which reached an intensity FWHM duration of 1.2 fs corresponding to 0.45 optical cycles at the initial central wavelength of 800 nm. The single-atom HHG spectrum for this driving waveform is shown in panel~\ref{fig:f2_timefreq}\textcolor{NavyBlue}{(e.1)} and visibly presents a lower cut-off energy than the spectrum corresponding to the previous 4-fs pulse. This is because the time-integrated electric field strength after the ionization step (i.e., the energy accumulated by the free electrons during their acceleration in the laser field) is smaller for the sub-cycle than for the few-cycle pulse. In addition, as we can see from the complete time-frequency analysis in Fig.~\ref{fig:f2_timefreq}\textcolor{NavyBlue}{(e)}, these sub-cycle IR fields tightly constrict the whole HHG process, with all the harmonics being emitted in a very narrow temporal window, leading to the direct generation of clean IAPs (see panel~\ref{fig:f2_timefreq}\textcolor{NavyBlue}{(e.2)}). Note that, despite their sub-cycle nature, the temporal duration of all the studied waveforms was found to be still sufficient to allow for the recombination HHG step to occur.

\subsection{Robustness of IAP generation against the CEP of sub-cycle drivers}
\label{sec:cep_robustness}

\begin{figure}[htbp]
    \centering
    \includegraphics[width=0.95\linewidth]{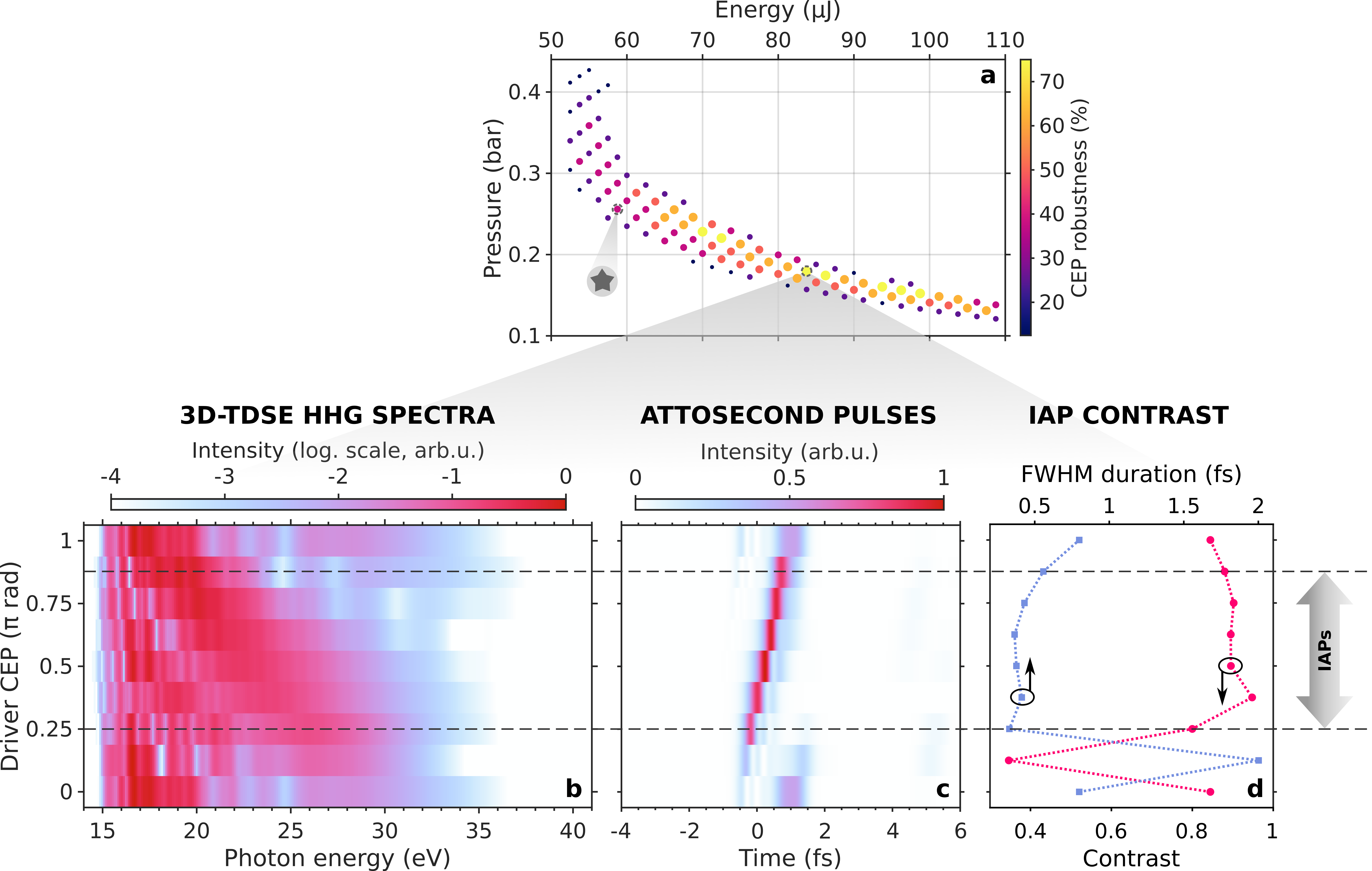}
    \caption{\textbf{(a)} Robustness of the IAPs to variations in the CEP of the sub-cycle drivers which are generated in the HCF for different pairs of pump energy and gas pressure. The highlighted point with a star label refers to the situation shown in Fig.~\ref{fig:f4_macroscopic}. \textbf{(b)} Single-atom HHG spectra as a function of the CEP of the IR driving field generated in the HCF for an input pulse energy $U_0=83.75$ \textmu J and an equivalent Ar pressure $p_{\mathrm{eq}}=179.5$ mbar. \textbf{(c)} Temporal profile and \textbf{(d)} FWHM intensity duration and contrast of the corresponding attosecond pulses.}
    \label{fig:f3_cep}
\end{figure}

We now focus on the situations where the self-compressed pulses from the HCF directly lead to the emission of high-order harmonics in the form of clean IAPs. In Fig.~\ref{fig:f3_cep}\textcolor{NavyBlue}{(a)} we plot all the investigated $(U_0,\,p_{\mathrm{eq}})$ pairs that produced IAPs. For a systematic search, we considered an attosecond pulse to be isolated whenever the intensity of the secondary temporal bursts remained below 10\% of the peak intensity. As we can clearly realize by comparing Figs.~\ref{fig:f2_timefreq}\textcolor{NavyBlue}{(a)} and \ref{fig:f3_cep}\textcolor{NavyBlue}{(a)}, the region for IAP generation matches the region for optimal sub-cycle self-compression in the HCF running along the contour line $L=L_{\mathrm{av}}$.

For these ultrashort IR driving waveforms, the intensity envelope $|A(\rho,z,T)|^2$ evolves in the same temporal scale as the carrier oscillations. Therefore, the CEP becomes of critical importance in the HHG process as it determines the particular variation of the laser electric field in time. Being driven by the electromagnetic field itself, all effects in strong-field laser interactions have demonstrated to be sensitive to the CEP \cite{Paulus2001,Zherebtsov2012}, and HHG is no exception. In particular, attosecond pulse production has been found to be very sensitive to the driver CEP when working with few-cycle pulses. In this regime, known as the non-adiabatic regime, CEP variations leave a clear imprint in the high-harmonic spectra due to the interference between consecutive attosecond bursts \cite{Baltuska2003,Nisoli2003,Borot2012,Rudawski2015,Hernandez-Garcia2015}, and adequate values of CEP can even lead to the generation of IAPs if the resulting electric field limits the emission of harmonics to a single recollision event. As a consequence, one would expect a similar or stronger dependence on the CEP for our self-compressed sub-cycle pulses.

To gain insight into these effects, we have first computed the robustness of the generated IAPs to variations in their driver CEP for each point in Fig.~\ref{fig:f3_cep}\textcolor{NavyBlue}{(a)}. This parameter is represented both in the color scale and with the markers size. For the sub-cycle pulse envelope generated in the HCF for each point $(U_0,\,p_{\mathrm{eq}})$, we defined the CEP robustness as the percentage of values of $\phi_{\mathrm{CEP}} \in [0,\pi]$ for which the resulting electric fields lead to the direct emission of IAPs according to the aforementioned isolation criterion. Noticeably, many cases exhibit very high values of CEP robustness around 50-60\%, the best cases reaching up to 75\%.

In addition, the complete 3D-TDSE HHG simulations results for an optimal case with 75\% of CEP robustness are shown in Figs.~\ref{fig:f3_cep}\textcolor{NavyBlue}{(b-d)}. From left to right, the three columns depict (b) the HHG spectra, (c) the corresponding temporal profile, and (d) the FWHM duration and contrast of the attosecond pulses generated with each sub-cycle IR driver as a function of its CEP. Here, contrast is defined as the ratio of energy within the main attosecond pulse to the overall energy transmitted through the aluminum filter. As we can see, clean IAPs are continuously generated in most of the CEP range. In all cases, these IAPs exhibit a great contrast above 0.8 and a FWHM duration of around 350-450 as. In the frequency domain, the generation of IAPs is accompanied by a smooth cut-off free of spectral fringes. Changes in the CEP of the driving field shift the recollision time along the pulse envelope \cite{Hernandez-Garcia2015}, resulting in a temporal drift of the attosecond pulses as clearly seen in Fig.~\ref{fig:f3_cep}\textcolor{NavyBlue}{(c)}. Despite these variations in emission times, HHG spectra and IAP durations, the most outstanding feature is that the single attosecond pulse isolation is preserved for most of the CEP range. This is presumably because, for optimally self-compressed IR drivers, the sub-cycle duration of the intensity envelope always limits the HHG process to a single recollision event from the only intense half cycle.

For those few cases where a second attosecond burst starts to show, the contrast drops abruptly and deep interference modulations appear in the harmonic spectra. It is also interesting to point out that these deteriorated cases appear around $\phi_{\mathrm{CEP}} = 0$, and not only the attosecond pulse is not isolated, but also the total yield is very low. This can be understood from the fact that two consecutive driving electric field peaks are involved in the HHG process, respectively for the ionization and recombination steps. In sub-cycle waveforms with $\mathrm{CEP} \sim 0$, which consist of a main peak surrounded by low-intensity structure, either the ionizing or recombinating field amplitude is weak, resulting in a low harmonic signal. On the contrary, sub-cycle drivers with intermediate values of CEP present two consecutive peaks with similar field strength and thus yield the brightest IAPs. Therefore, sub-cycle pulses with $\phi_{\mathrm{CEP}} = 0$ appear not to be the ideal drivers for HHG, and we expect even shorter pulses with these CEP values to prevent any EUV emission due to the absence of returning field to drive the free electrons back to the parent ion \cite{Hassan2016}.

\subsection{IAP robustness against macroscopic HHG}

Previous results for the whole set of IR waveforms were obtained from single-atom 3D-TDSE calculations, already providing a general insight into the properties of HHG driven by self-compressed pulses from HCFs. However, being a highly nonlinear process, the amplitude and phase of the harmonic emission is very sensitive to the details of the driving field. As a result, a complete description of HHG should include propagation and phase-matching of the harmonics in the gas target \cite{Salieres1995,Rundquist1998,Gaarde2008,Popmintchev2009,Hernandez-Garcia2010,Hernandez-Garcia2013,Weissenbilder2022}. In particular for ultra-broadband sub-cycle waveforms, diffraction-induced spatio-temporal reshaping and changes in the CEP of the driving pulse around the focal volume can affect the efficient build-up of the harmonics emitted across the target and hinder the isolation of clean attosecond pulses.

\begin{figure}[htbp]
    \centering
    \includegraphics[width=0.95\linewidth]{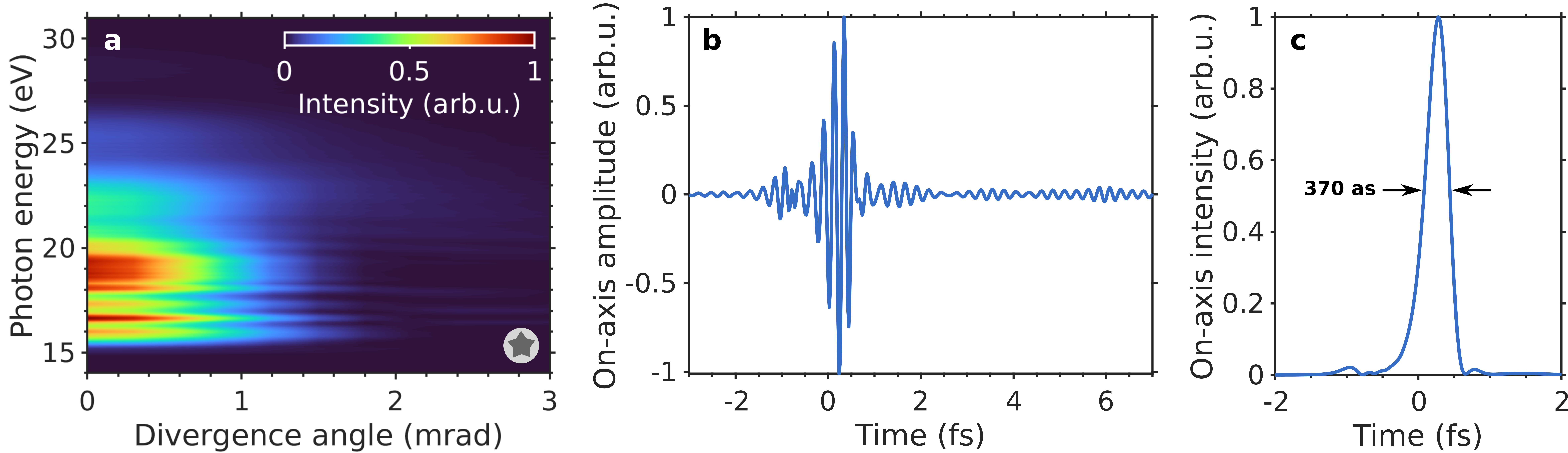}
    \caption{\textbf{(a)} Harmonic spectrum as a function of the divergence angle in the far-field detector obtained by driving the HHG process, in a 1-mm-thick low-density hydrogen jet centered at the beam focus, with the sub-cycle IR pulse obtained from the HCF for $U_0=58.75$ \textmu J, $p_{\mathrm{eq}}=255.8$ mbar and $\phi_{\mathrm{CEP}}=5\pi/8$ rad (corresponding to the situation labeled in Fig.~\ref{fig:f3_cep}\textcolor{NavyBlue}{(a)} with a star) and focused and attenuated to an instantaneous peak intensity of $1.82 \times 10^{14}$ W/cm$^2$ (field amplitude of 0.072 au). \textbf{(b)} Resulting on-axis attosecond pulse and \textbf{(c)} its temporal intensity profile.}
    \label{fig:f4_macroscopic}
\end{figure}

To analyze the impact of macroscopic propagation in the generation of IAPs, we have performed a complete HHG simulation in a gas jet for $U_0=58.75$ \textmu J and $p_{\mathrm{eq}}=255.8$ mbar. According to the single-atom calculations, this situation presented a low CEP robustness of $\sim$37\% as shown in Fig.~\ref{fig:f3_cep}\textcolor{NavyBlue}{(a)} with the star label. This relatively sensitive case was chosen to study the influence of propagation because we expect the optimal cases presenting higher CEP robustness to be less affected by phase-matching. Using these input parameters, we computed the (2+1)D self-compression of the pump pulse in the HCF. Although the output on-axis temporal profile was found to be almost identical to the one previously obtained from the (1+1)D model, the (2+1)D simulation provided the real spatial profile at the fiber end. This beam was then freely propagated in vacuum along a distance of 1 m and subsequently focused with one spherical mirror ($f = 30$ cm) to a spot radius of approximately 30 \textmu m, corresponding to a Rayleigh length $z_R \sim 3.5$ mm at 800 nm. The driving waveform was built with $\phi_{\mathrm{CEP}} = 5\pi/8$ rad, as this yielded the cleanest IAP in the single-atom simulations, and attenuated to an instantaneous peak intensity of $1.82 \times 10^{14}$ W/cm$^2$ to avoid barrier suppression. The complete IR pulse around the focal volume was finally used as input for macroscopic HHG calculations in a 1-mm-thick, low-density hydrogen target centered at the beam focus. Figure~\ref{fig:f4_macroscopic}\textcolor{NavyBlue}{(a)} shows the resulting harmonic spectrum as a function of the divergence angle in the far-field detector, after spatial integration along the azimuthal coordinate. All harmonics are emitted with a low divergence $<1$ mrad, and a highly contrasted 370-as IAP is generated across the whole EUV beam. Its on-axis temporal amplitude and intensity profile are shown in Figs.~\ref{fig:f4_macroscopic}\textcolor{NavyBlue}{(b,c)}. This is in good agreement with the predictions of the single-atom 3D-TDSE calculations, and we expect the optimal cases exhibiting the highest CEP robustness to also tolerate phase-matching effects, as further confirmed by additional macroscopic simulations in longitudinal and transversal targets not shown here. These results demonstrate that the proposed scheme is indeed capable of generating high-quality IAPs which remain stable upon CEP variations and phase-matching in a gas target. Although it is out of the scope of this paper, further macroscopic HHG optimization by controlling beam size, wavefront curvature, and gas jet pressure and relative position to beam focus, could be additionally used to modify the IAP properties, such as pulse duration, chirp or divergence \cite{Salieres1995,Hernandez-Garcia2013,Wikmark2019,Quintard2019,Rego2022,Weissenbilder2022}.

\subsection{High-energy sub-cycle self-compression towards an all-fiber IAP source}

\begin{figure}[htbp]
    \centering
    \includegraphics[width=0.95\linewidth]{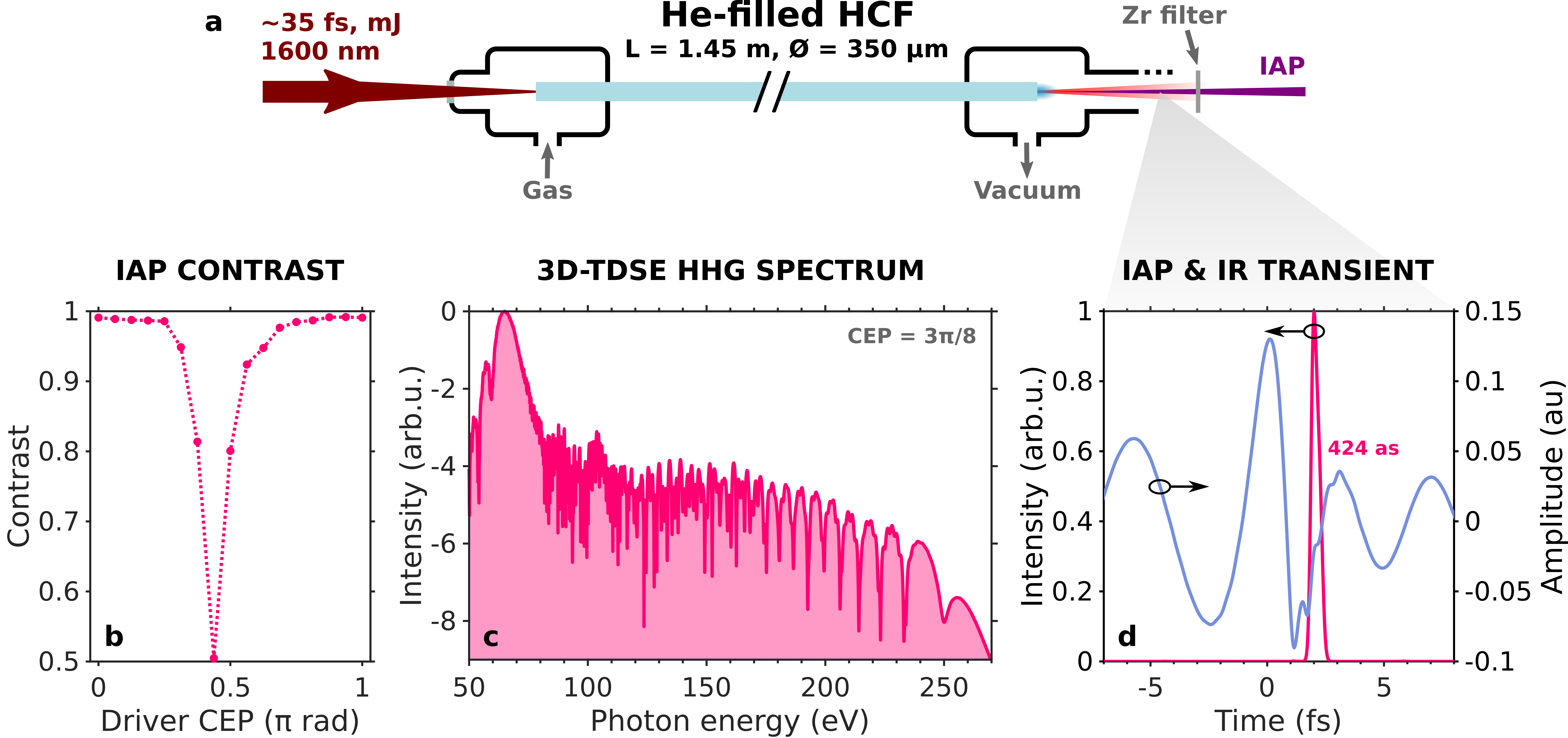}
    \caption{\textbf{(a)} Schematic of an all-fiber IAP source, where HHG is directly driven at the HCF end by a self-compressed, high-energy IR sub-cycle transient. \textbf{(b)} Contrast of the directly emitted IAPs as a function of the CEP of the driving 2.2-fs, 1.3-mJ self-compressed pulse at 1600 nm, generated in the helium-filled HCF. \textbf{(c)} Single-atom 3D-TDSE high-harmonic spectrum in He for the waveform with $\phi_{\mathrm{CEP}}=3\pi/8$ rad, after transmission through a zirconium filter. \textbf{(d)} Corresponding IAP (fuchsia line) plotted over the main feature of the IR driving field.}
    \label{fig:f5_allfiber}
\end{figure}

Given the discussion above on IAP generation following extreme pulse self-compression in a single HCF step, it is natural to ask whether the scheme presented in Fig.~\ref{fig:f1_setup} could be further simplified by removing the focusing stage and generating high-order harmonics directly at the fiber end, as shown in Fig.~\ref{fig:f5_allfiber}\textcolor{NavyBlue}{(a)}. All-fiber HHG sources have been briefly envisioned both theoretically and experimentally in hollow-core photonic crystal fibers \cite{Travers2011,Serebryannikov2008,Heckl2009}, and a compact soft X-ray source, enabled by self-compression of $\sim 2$-\textmu m pulses in an anti-resonant HCF, has been recently demonstrated \cite{Gebhardt2021}. Nevertheless, to the best of our knowledge, this scenario has not yet been explored in simpler and larger-core HCFs, where high intensities can be routinely reached. In these systems, achieving a sufficiently high intensity at the fiber output so as to directly drive efficient HHG would require pumping with millijoule-level pulses and carefully choosing a relatively small core radius to boost the peak intensity without causing unaffordable losses. However, at high intensities, ionization might also become a problem, both because it leads to severe distortion of the self-compressing pulse and because it complicates the phase-matching of the harmonics due to strong free-electron dispersion. Furthermore, as a high energy pulse self-compresses towards sub-cycle durations, its peak power drastically increases, which can lead to beam self-focusing and additional gas ionization if the peak power exceeds the critical value $P_{\mathrm{cr}}$ \cite{Crego2019}. Fortunately, as both ionization and $P_{\mathrm{cr}}$ scale inversely with the gas density, these detrimental nonlinear effects can be ameliorated with the decreasing pressure gradient configuration.

As a proof of concept of an all-fiber IAP source, we have followed the scaling rules presented in Section \ref{sec:scaling} to translate sub-cycle self-compression to millijoule-level pulses which could directly drive HHG at the fiber end. One immediately realizes that, to achieve a high-quality compression, the constraint $N<15$ inevitably sets an upper limit to the maximum pump energy. By further inspection of the soliton order, it is straightforward to prove that, in a first approximation where only the waveguide dispersion is considered, $N \propto (p_{\mathrm{eq}}T_pU_0/\lambda_0^4)^{1/2}$. As a result, if $N$ is to take some optimal value below 15, then $U_0 \propto N^2\lambda_0^4/(p_{\mathrm{eq}}T_p)$ and, thus, the most efficient way to upscale the pulse energy is to increase the initial central wavelength $\lambda_0$ by moving towards the mid-IR spectral region. The combination of higher energies and longer wavelengths would definitely be of great interest for HHG experiments. Indeed, short-wavelength- and mid-IR drivers have been widely used to achieve high photon energies up to the soft X-rays \cite{Chen2010,Popmintchev2012,Teichmann2016,Johnson2018}. Now, thanks to a stronger HCF anomalous response at longer wavelengths, this energy scaling can be accomplished even in more practical fibers ($L\sim$1 m).

As an example of this scaling principle, we have used the complete (2+1)D model to simulate the self-compression of an input 2-mJ, 35-fs gaussian pulse centered at 1600 nm through a 1.45-m-long, 175-\textmu m-core-radius, helium-filled HCF with a decreasing pressure gradient from $p_0=3.15$ bar to $p_L=0.15$ bar. In this case, He was chosen to avoid problems with ionization and the fiber end was left with some pressure instead of completely evacuated so that the HHG generation medium could be the filling gas itself. Nevertheless, we verified with the simulations that the output pressure $p_L$ could be approximately varied in the range from 0 to 0.3 bar without incurring any noticeable distortions to the self-compressed driver. This fact could give some freedom to adjust the phase-matching of the high-harmonics. In the configuration with $p_L=0.15$ bar, the IR pulse self-compresses to an output sub-cycle duration of 2.2 fs, corresponding to 0.41 optical cycles at 1600 nm, and retains 1.3 mJ of energy. This yields a peak intensity of $7.2\times 10^{14}$ W/cm$^2$ at the fiber end, or a peak power of 0.2 TW when integrating over the spatial profile, which is high enough to directly drive HHG in He.

After adding the carrier wave with different values of $\phi_{\mathrm{CEP}}$, the resulting waveforms were used to perform single-atom 3D-TDSE HHG calculations in He. Note that full macroscopic HHG simulations at this longer IR wavelength using the 3D-TDSE are beyond the state-of-the-art computational capabilities: first because the non-adiabatic nature of the sub-cycle pulses precludes the use of faster algorithms like those relying on the strong-field approximation and instead requires the use of the full-quantum 3D-TDSE \cite{Becker2001}, and second because its numerical integration becomes very demanding already at the microscopic level, with high photon energies requiring a small time step and long wavelengths requiring a large spatial grid to fit the electron trajectories. Here, as opposed to previous simulations, the driving electric fields were directly used with the amplitude corresponding to the 1.3-mJ pulse at the HCF end for each CEP, without applying any normalization. Reaching higher photon energies in He than in hydrogen, the HHG spectra were now filtered with a 200-nm-thick zirconium foil. As we can see in Fig.~\ref{fig:f5_allfiber}\textcolor{NavyBlue}{(b)}, this configuration also yields IAPs with a very high contrast approaching unity for more than half of the complete CEP range. For instance, Fig.~\ref{fig:f5_allfiber}\textcolor{NavyBlue}{(c)} shows the broadband HHG spectrum generated by the waveform with $\phi_{\mathrm{
CEP}}=3\pi/8$ rad, which in the temporal domain yields the clean IAP plotted in Fig.~\ref{fig:f5_allfiber}\textcolor{NavyBlue}{(d)} over the main feature of the corresponding IR sub-cycle field.

%%%----------------------------------------------------------------------------
\section{Conclusions}

In conclusion, we have theoretically demonstrated a compact and robust scheme for generating EUV IAPs from high-order harmonics. Starting from a standard multi-cycle IR pulse, a light transient is generated by extreme soliton self-compression in a negatively pumped HCF, and is subsequently used to drive HHG in a gas target leading to the direct emission of IAPs without the need for additional gating techniques. Systematic nonlinear pulse propagation simulations combined with full-quantum 3D-TDSE HHG calculations have shown that high-contrast IAPs are directly emitted for a broad set of driving fields corresponding to the optimally self-compressed IR pulses. This provides a general route towards robust IAP generation for any HCF configuration, since the pump energy and gas pressure which lead to optimal sub-cycle pulse compression, and thus to IAP emission, can be identified in a universal manner by matching the fiber length to an average self-compression length. Most remarkably, owing to the nature of the IR waveforms, the single attosecond pulse isolation is preserved for most of the driver CEPs, presumably because the sub-cycle duration of the intensity envelope continuously constricts the HHG process to a single recollision event from the only intense half cycle of the electric field. In addition to being CEP robust, the proposed scheme has also shown to be stable under macroscopic propagation including phase-matching of the high-order harmonics in a low-density gas target. Finally, we have provided preliminary theoretical advice for the development of all-fiber IAP sources driven by self-compressed millijoule-level sub-cycle IR pulses. Altogether, we believe that these findings might pave the way towards a new generation of compact and robust experiments for IAP generation with sub-cycle drivers, which, among other applications, offer great promise for advancing real-time observation and precision control of electron dynamics at the atomic scale.

%%%----------------------------------------------------------------------------
\section*{Acknowledgments}

This project has received funding from Ministerio de Ciencia e Innovación (MCIN/AEI/10.13 039/501100011033, I+D+i grants PID2019-106910GB-I00 and PID2022-142340NB-I00) and from the European Research Council (ERC) under the European Union's Horizon 2020 research and innovation program (Grant Agreements No. 851201 and No. 951224). M.F.G. acknowledges support from Ministerio de Universidades under grant FPU21/02916. The authors thankfully acknowledge the computer resources at MareNostrum and SCAYLE, and the technical support provided by Barcelona Supercomputing Center (RES-FI-2022-3-0041).

\end{document}